\documentclass{nature}

\usepackage{amsmath}
\usepackage{amssymb}
\usepackage{mathrsfs}
\usepackage{paralist}
\usepackage{hyperref}
\usepackage{booktabs}
\usepackage{graphicx}
\usepackage{float}
\usepackage{morefloats}
\usepackage{multirow}
\usepackage{color}
\usepackage{paralist}
\usepackage{todonotes}
\usepackage{wrapfig}
\usepackage{bbding}

\newcommand\blfootnote[1] 
{
  \begingroup
  
  \footnotetext{#1}
  \addtocounter{footnote}{-1}
  \endgroup
}

\title{Learning MRI Artifact Removal With Unpaired Data}

\author{Siyuan Liu\textsuperscript{1}, Kim-Han Thung\textsuperscript{1}, Liangqiong Qu\textsuperscript{1}, Weili Lin\textsuperscript{1}, 
  Dinggang~Shen\textsuperscript{1}, 
  Pew-Thian Yap\textsuperscript{1,\Envelope}, and the UNC/UMN Baby Connectome Project Consortium}
\begin{document}

\maketitle

\begin{affiliations}
 \item Department of Radiology and Biomedical Research Imaging Center (BRIC), University of North Carolina at Chapel Hill, NC, U.S.A.
\end{affiliations}

\begin{abstract}
Retrospective artifact correction (RAC) improves image quality post acquisition and enhances image usability. Recent  machine learning driven techniques for RAC are predominantly based on supervised learning and therefore practical utility can be limited as data with paired artifact-free and artifact-corrupted images are typically insufficient or even non-existent. 
Here we show that unwanted image artifacts can be disentangled and removed from an image via an RAC neural network learned with unpaired data. This implies that our method does not require matching artifact-corrupted data to be either collected via acquisition or generated via simulation.
Experimental results demonstrate that our method is remarkably effective in removing artifacts and retaining anatomical details in images with different contrasts.
\end{abstract}

\blfootnote{\noindent \Envelope~Corresponding author: Pew-Thian Yap (\texttt{ptyap@med.unc.edu})}

\section*{Introduction}
Structural magnetic resonance imaging (sMRI) captures high spatial-resolution details of brain anatomy, but is susceptible to artifacts caused for example by eye and head motions\cite{Budde2014Ultra}, especially when scanning pediatric, elderly, claustrophobic, and epileptic patients\cite{Zhuo2006MR}.
Artifacts can result in unusable images and hence cause financial losses for imaging studies\cite{Andre2015Toward}.
Motion artifact correction\cite{Zaitsev2015Motion} can be used to remove artifacts, improve image quality, and increase the amount of usable images. 
This is particularly important in view of the fact that the accuracy and reliability of subsequent analysis or diagnosis can be jeopardized by poor image quality.

Methods for correction of motion artifacts can be prospective or retrospective. 
Prospective techniques\cite{Zaitsev2006Magnetic,Qin2009Prospective,Ooi2009Prospective,Schulz2012An,Maclaren2012Measurement, Maclaren2012Prospective} utilize either optical tracking of target markers placed on the head or continuously reacquired images from dedicated navigator scans 
for real-time motion prediction\cite{Schulz2012An}. 
However, 
prospective methods require additional hardware and/or scanner modifications. Motion markers can cause patient discomfort and 
optical tracking requires expensive hardware, needs clear visibility of target markers, and may be sensitive to facial movements.
Moreover, these methods typically assume quiescent periods with minimal motion and reacquire data when this condition is not met.
This prolongs acquisition time without necessarily bringing substantial improvements to image quality when there is little motion.

In contrast, retrospective artifact correction (RAC)\cite{Zaitsev2015Motion} can be used to remove artifacts, improve image quality, and enhance image usability without requiring additional hardware, as motion estimation and correction are considered a part of the image reconstruction process. 
Retrospective techniques can be acquisition-based or software-based. 
One representative technique of acquisition-based methods is PROPELLER\cite{Pipe1999Motion}, which is a self-navigation technique that
utilizes a number of concentric blades rotated at different angles to cover the whole k-space. The k-space center is repeatedly sampled and used as self-navigators for the estimation of rotation and translation information. It has been shown to be effective for 2D motion correction\cite{Vertinsky2009Performance}.
However, acquisition-based methods such as PROPELLER can only estimate in-plane motion parameters, and the through-plane motion might disrupt signals across slices.
Moreover, 
retrospective techniques often require additional purposefully designed navigator sequences, more complicatedly designs,
longer acquisition times, and impose additional constraints on imaging parameters (i.e., TR/TE/TI). 

Software-based methods for post-acquisition RAC\cite{Zaitsev2015Motion} can be used to remove artifacts
without modification of sequences, mounting of markers, and constraining acquisition parameters. They are inexpensive post-processing method that can be readily incorporated across all scanners.
Particularly, deep neural networks (DNNs), such as convolutional neural networks (CNNs), have demonstrated great potential for simultaneous removal of a variety of artifacts irrespective of the acquisition scheme\cite{Jin2017Deep,Haskell2019Network}.
CNNs are typically trained in a supervised manner, which in RAC requires paired artifact-corrupted and artifact-free images.
Such paired data can be collected by scanning the same subjects without and with motions, which can be impractical, costly, and time-consuming.
Artifact-corrupted images can also be generated by adding simulated artifacts to artifact-free images \cite{Patricia2018Motion,Tamada2019Motion,Kuestner2019Retrospective,Johnson2019Conditional}. 
However, simulations might not accurately and sufficiently reflect all possible forms of real artifacts. 

In this paper, we consider the artifact removal problem as image translation from an artifact-corrupted domain to an artifact-free domain.
We gain inspirations from unsupervised image translation techniques, such as UNIT\cite{Liu2017Unsupervised}, CycleGAN\cite{Zhu2017Unpaired}, BicycleGAN\cite{Zhu2017Toward}, and Pix2Pix\cite{Isola2017Image}, which employ auto-encoders to learn invertible mappings between domain pairs using unpaired images.
We introduce an end-to-end disentangled unsupervised cycle-consistent adversarial network (DUNCAN), which can be trained using unpaired data for flexible and simultaneous removal of various sMRI artifacts.
We employ cycle translations between artifact-corrupted and artifact-free domains, where each cycle translation is defined as a forward translation from one domain to its target domain, followed by a backward translation from the target domain to the original domain. 
Both the forward and backward translations are realized with auto-encoders.
Note that each MR image, even deemed good in quality, may inevitably contain some artifacts. Therefore, we assume that images from either artifact-corrupted or artifact-free domains are composed of an anatomical content component, residing in a domain-invariant content space, and an artifact component, residing in a domain-specific artifact space.
The auto-encoders disentangle these two components in an image via two kinds of encoders in each domain translation mapping, i.e., a content encoder, which captures anatomical structures shared across domains, and an artifact encoder, which captures artifacts specific to a domain.
Then, the decoder ensembles the extracted content and artifact features from both encoders to translate images to the target domain.
To ensure complete disentanglement of content and artifact components, we propose a multi-scale content consistency (MS-CC) loss and a content-swapping mechanism supervised by adversarial learning.
We also design a multi-scale reconstruction consistency (MS-RC) loss, including a pixel reconstruction consistency (PRC) loss, an edge reconstruction consistency (ERC) loss, and a structure reconstruction consistency (SRC) loss, to avoid degradation of structural details.
In addition, we propose an image quality consistency (IQC) loss to ensure that no structural details are removed from artifact-free images.
The architecture of DUNCAN is summarized in Figure 1 and detailed in the Methods section.
\begin{figure*}[!t]
	\centering
	\includegraphics[width=\textwidth]{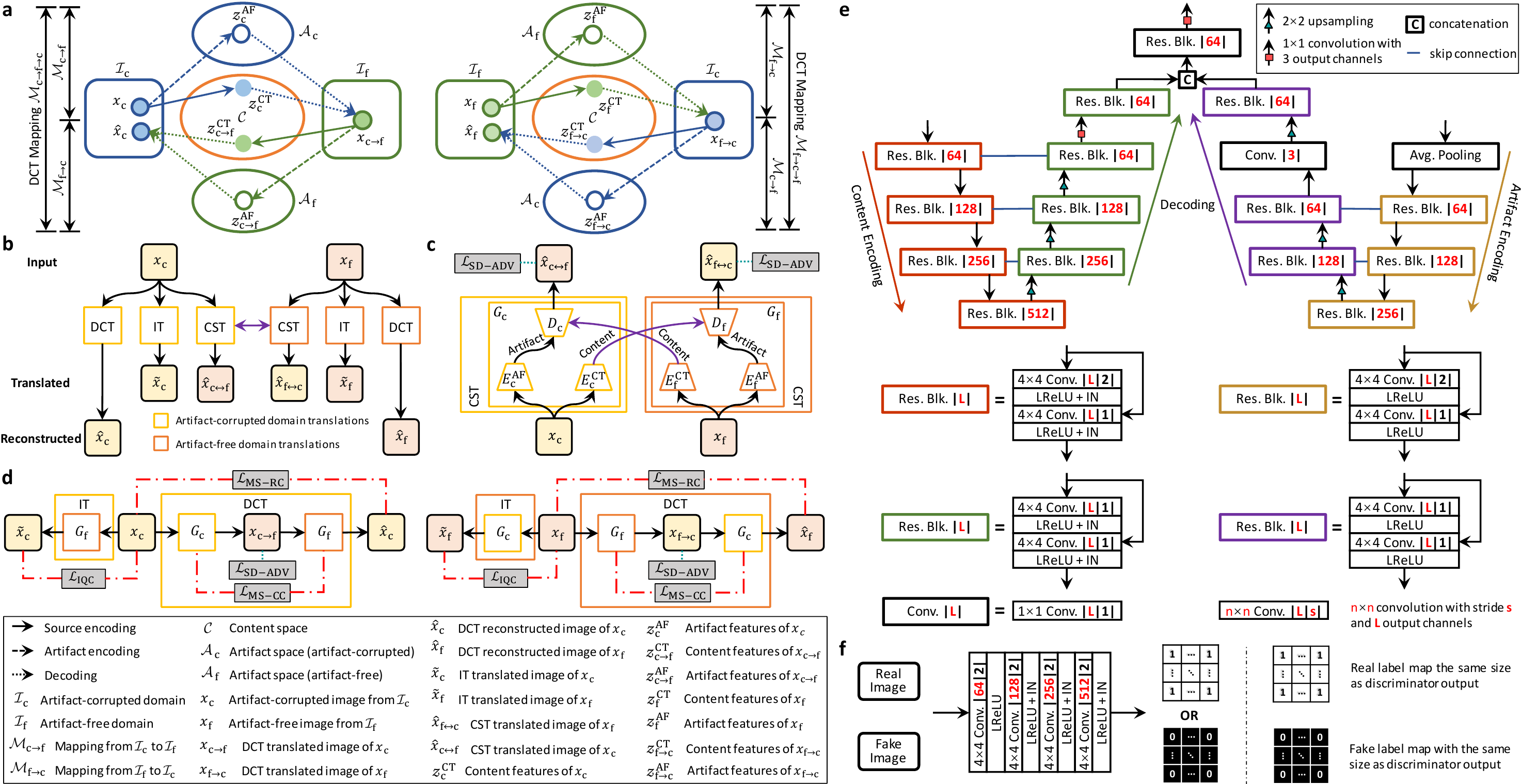}
	\vspace{-10pt}
	\caption{\textbf{Overview of DUNCAN.} \textbf{a}, Disentangled cycle translation (DCT) mapping $\mathcal{M}_{\text{c}\rightarrow \text{f}\rightarrow \text{c}}$ consists of two sequential domain mappings $\mathcal{M}_{\text{c}\rightarrow \text{f}}$ and $\mathcal{M}_{\text{f}\rightarrow \text{c}}$. For $\mathcal{M}_{\text{c}\rightarrow \text{f}}$, an artifact-corrupted image $x_{\text{c}}$ is first encoded in the domain-invariant content space $\mathcal{C}$ and the domain-specific artifact space $\mathcal{A}_{\text{c}}$ to obtain the content (CT) information $z_{\text{c}}^{\text{CT}}$ and artifact (AF) information $z_{\text{c}}^{\text{AF}}$, respectively. Then, $z_{\text{c}}^{\text{CT}}$ and $z_{\text{c}}^{\text{AF}}$ are decoded to remove artifacts from $x_{\text{c}}$, and to obtain the intermediate translated image $x_{\text{c}\rightarrow \text{f}}$ in the artifact-free domain $\mathcal{I}_{\text{f}}$. For $\mathcal{M}_{\text{f}\rightarrow \text{c}}$, $x_{\text{c}\rightarrow \text{f}}$ is first encoded in the content space $\mathcal{C}$ and the artifact space $\mathcal{A}_{\text{f}}$ to obtain the content information $z_{\text{f}}^{\text{CT}}$ and the artifact information $z_{\text{f}}^{\text{AF}}$, respectively. Then $z_{\text{f}}^{\text{CT}}$ and $z_{\text{f}}^{\text{AF}}$ are decoded to add artifacts to $x_{\text{c}\rightarrow \text{f}}$ to reconstruct image $\hat{x}_{\text{c}}$.
		DCT mapping $\mathcal{M}_{\text{c}\rightarrow \text{f}\rightarrow \text{c}}$ is hence $\{x_{\text{c}} \in \mathcal{I}_{\text{c}}\} \rightarrow \{z_{\text{c}}^{\text{CT}} \in \mathcal{C}, z_{\text{c}}^{\text{AF}} \in \mathcal{A}_{\text{c}} \} \rightarrow \{x_{\text{c}\rightarrow \text{f}} \in \mathcal{I}_{\text{f}}\} \rightarrow \{z_{\text{c}\rightarrow \text{f}}^{\text{CT}} \in \mathcal{C}, z_{\text{c}\rightarrow \text{f}}^{\text{AF}} \in \mathcal{A}_{\text{f}} \} \rightarrow \{\hat{x}_{\text{c}} \in \mathcal{I}_{\text{c}}\}$.
		Conversely, 
		DCT mapping $\mathcal{M}_{\text{f}\rightarrow \text{c}\rightarrow \text{f}}$ is $\{x_{\text{f}} \in \mathcal{I}_{\text{f}}\} \rightarrow \{z_{\text{f}}^{\text{CT}} \in \mathcal{C}, z_{\text{f}}^{\text{AF}} \in \mathcal{A}_{\text{f}} \} \rightarrow \{x_{\text{f}\rightarrow \text{c}} \in \mathcal{I}_{\text{c}}\} \rightarrow \{z_{\text{f}\rightarrow \text{c}}^{\text{CT}} \in \mathcal{C}, z_{\text{f}\rightarrow \text{c}}^{\text{AF}} \in \mathcal{A}_{\text{c}} \} \rightarrow \{\hat{x}_{\text{f}} \in \mathcal{I}_{\text{f}}\}$.
		\textbf{b}, DUNCAN takes any two unpaired images, i.e., one image $x_{\text{c}}$ from artifact-corrupted domain $\mathcal{I}_{\text{c}}$ and one image $x_{\text{f}}$ from artifact-free domain $\mathcal{I}_{\text{f}}$, as inputs.
		The artifact-corrupted and artifact-free domain cycles incorporate the DCT mappings $\mathcal{M}_{\text{c}\rightarrow \text{f}\rightarrow \text{c}}$ and $\mathcal{M}_{\text{f}\rightarrow \text{c}\rightarrow \text{f}}$, respectively.
		The content-swapping translation (CST) and identity translation (IT), respectively, give the content-swapped translated images, i.e., $\hat{x}_{\text{f} \leftrightarrow \text{c}}$ and $\hat{x}_{\text{c} \leftrightarrow \text{f}}$, and identity translated images, i.e., $\tilde{x}_{\text{c}}$ and $\tilde{x}_{\text{f}}$.
		\textbf{c}, CST in artifact-corrupted and artifact-free domains.
		\textbf{d}, DCT and IT in artifact-corrupted and artifact-free domains.
		\textbf{e}, Network architecture of the proposed auto-encoder ($G_{\text{c}}$ or $G_{\text{f}}$).
		\textbf{f}, Network architecture of the discriminator. The discriminator employs a fully convolutional network (FCN) to determine if the generate image is real or fake based on a semantic map\cite{Isola2017Image}.
	}
	\label{fig:framework}
\end{figure*}
\section*{Results}
\begin{figure*}[!t]
	\centering
	\includegraphics[width=0.99\textwidth]{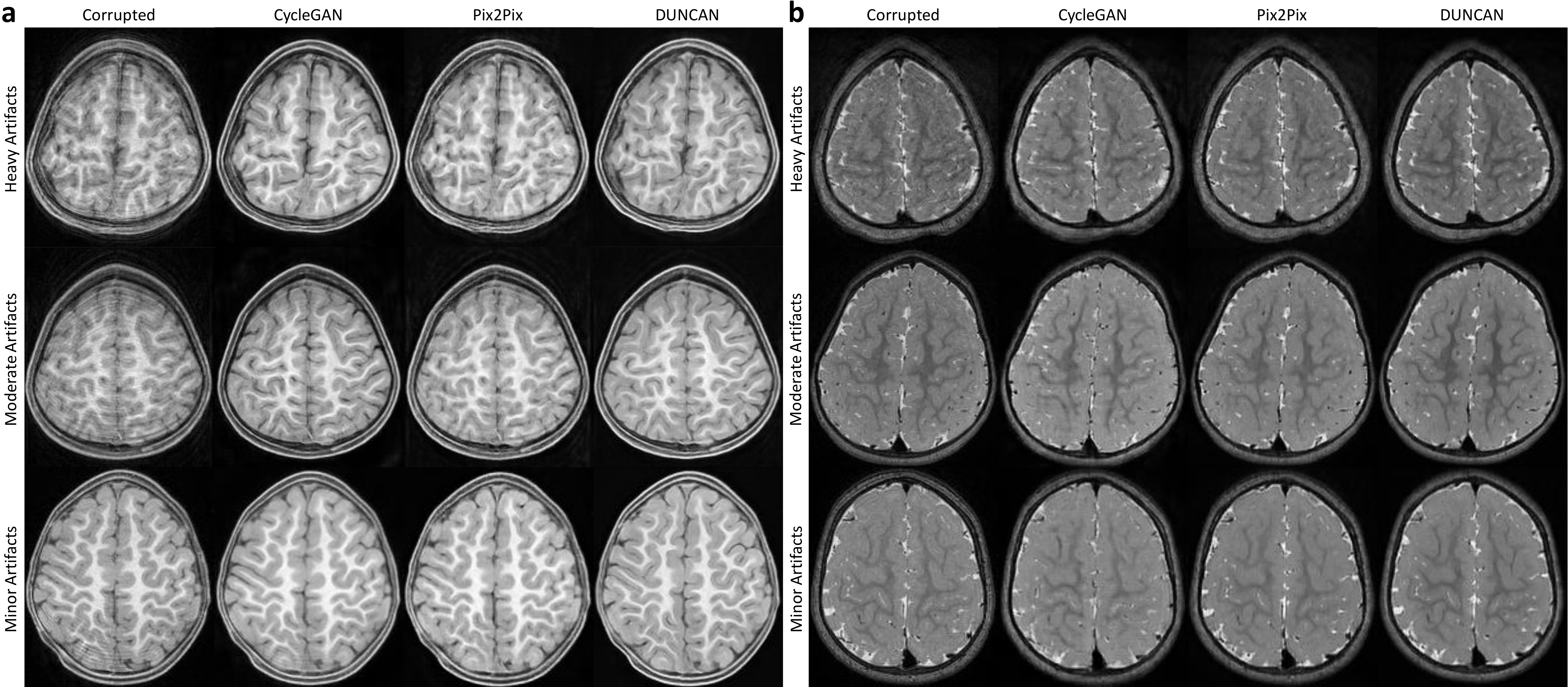}
\caption{\textbf{Visual comparison of corrected in vivo images.} 
	\textbf{a}, T1-weighted images and, \textbf{b}, T2-weighted images corrected using various methods.
	From top to bottom are images with heavy, moderate, and minor artifacts. 
	In \textbf{a} and \textbf{b}, the original artifact-corrupted images are shown in the first column and the images corrected using CycleGAN, Pix2Pix, and DUNCAN are shown respectively in the second to fourth columns.
	DUNCAN outperforms the other methods in removing artifacts and in preserving anatomical details.}
\label{fig:CL_comp}
\end{figure*}
\begin{figure*}[!t]
	\centering
	\includegraphics[width=0.99\textwidth]{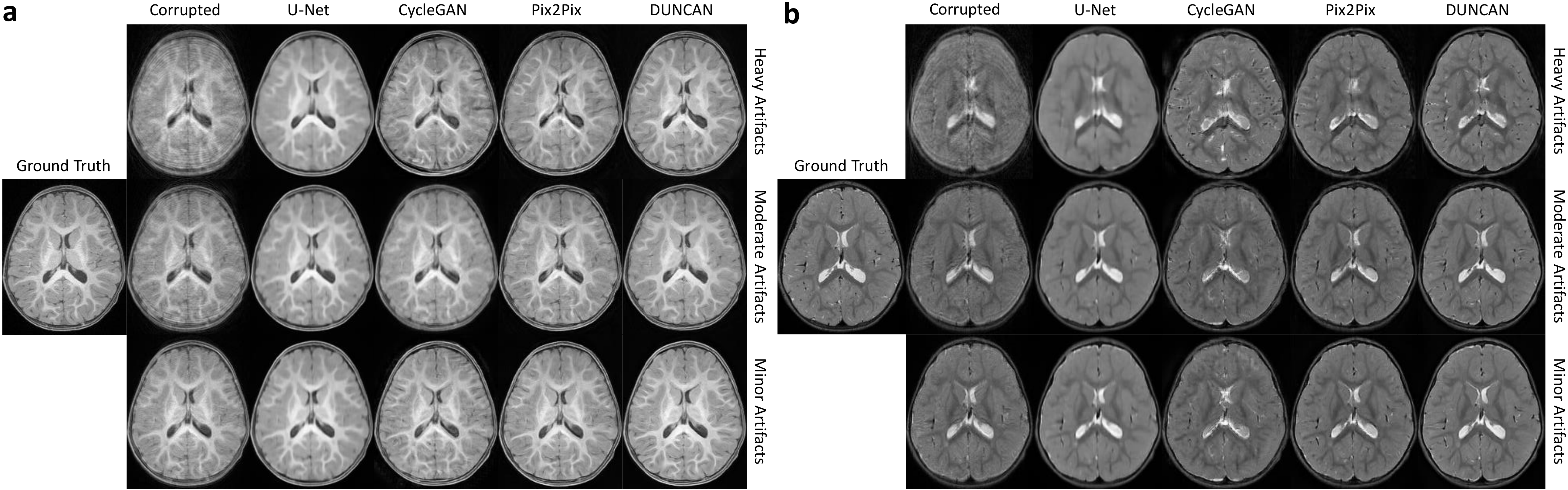}
	\caption{\textbf{Visual comparison of corrected in silico images.} 
		\textbf{a}, T1-weighted images and, \textbf{b}, T2-weighted images corrected using various methods.
  From top to bottom are images with heavy, moderate, and minor artifacts. 
	In \textbf{a} and \textbf{b}, the ground truth is shown in the first column, the original artifact-corrupted images in the second column, and the images corrected using U-Net, CycleGAN, Pix2Pix, and DUNCAN, respectively, in the third to sixth columns. DUNCAN removes more artifacts and preserves more anatomical details in agreement with the ground truth.}
	\label{fig:SYN_comp_qual}
\end{figure*}
\begin{figure*}[!t]
	\centering
	\includegraphics[width=\textwidth]{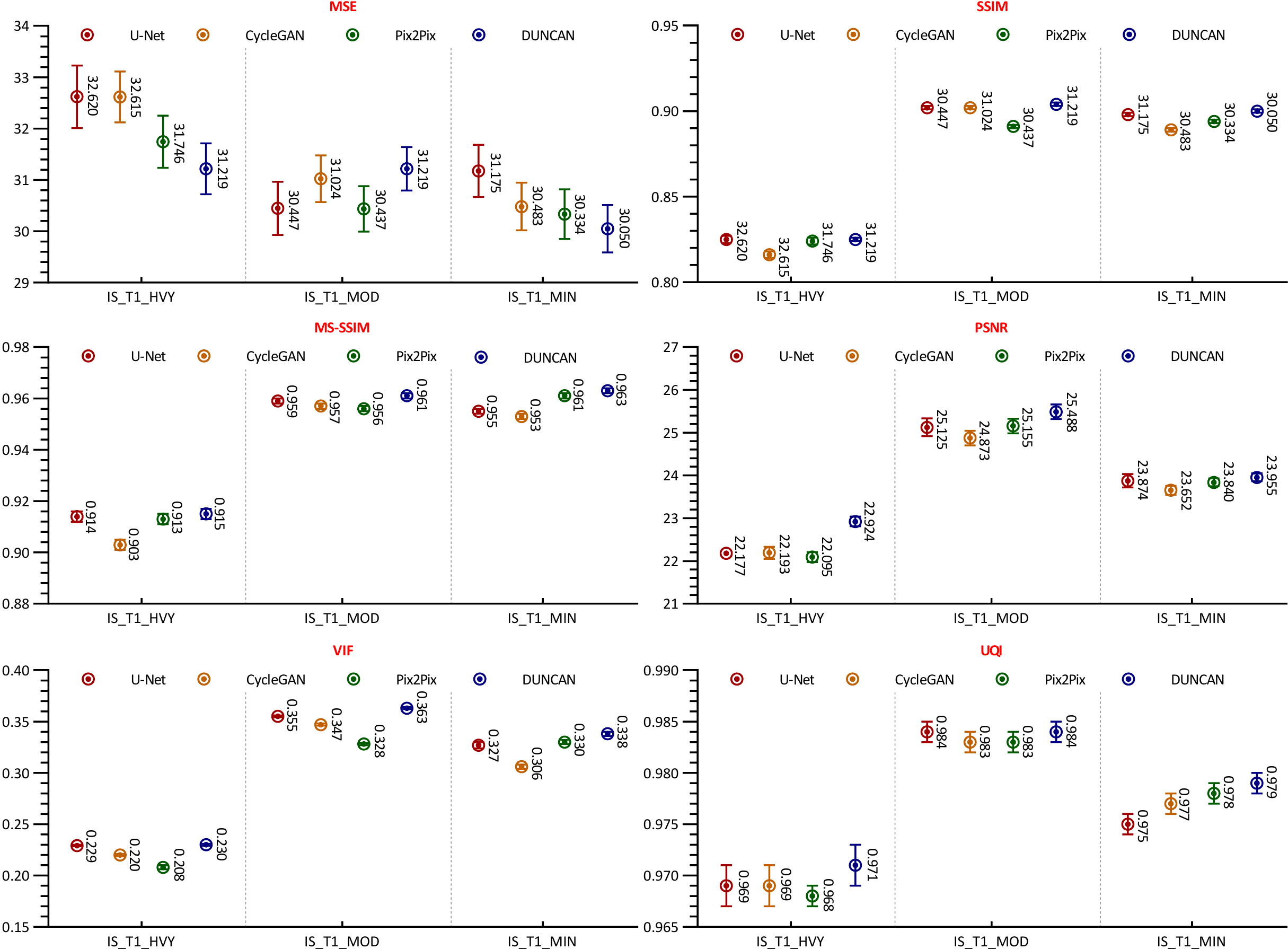}
	\vspace{-20pt}
	\caption{\textbf{Quantitative comparison of corrected in silico T1-weighted images.} Numerical evaluation conducted with different levels of artifacts (heavy, moderate, and minor) and various metrics (MSE, SSIM, MS-SSIM, PSNR, VIF, and UQI). The bars show the means and the error bars show the standard errors on the means. The sample sizes of IS\_T1\_HVY, IS\_T1\_MOD, and IS\_T1\_MIN are 200, 300, 300, respectively. Compared with the other methods, DUNCAN yields lower MSE and higher SSIM, MS-SSIM, PSNR, VIF, and UQI.}
	\label{fig:SYN_T1_comp_quan}
\end{figure*}
\begin{figure*}[!t]
	\centering
	\includegraphics[width=\textwidth]{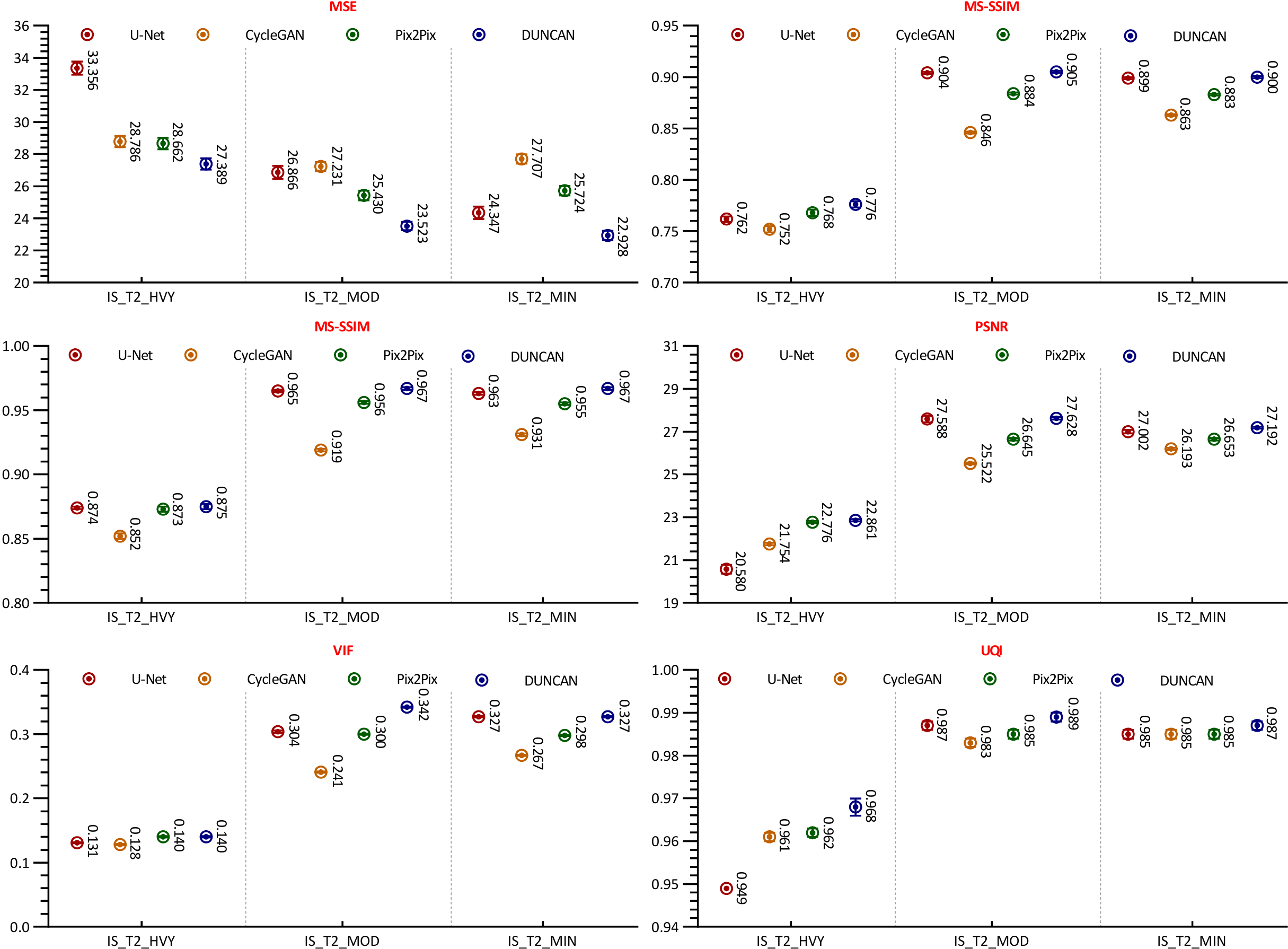}
	\vspace{-20pt}
	\caption{\textbf{Quantitative comparison of corrected in silico T2-weighted images.} Numerical evaluation conducted with different levels of artifacts (heavy, moderate, and minor) and various metrics (MSE, SSIM, MS-SSIM, PSNR, VIF, and UQI). The bars show the means and the error bars show the standard errors on the means. The sample sizes of IS\_T2\_HVY, IS\_T2\_MOD, and IS\_T2\_MIN are 200, 300, 300, respectively. Compared with the other methods,  DUNCAN yields lower MSE and higher SSIM, MS-SSIM, PSNR, VIF, and UQI.}
	\label{fig:SYN_T2_comp_quan}
\end{figure*}
\begin{figure*}[!t]
	\centering
	\includegraphics[width=\textwidth]{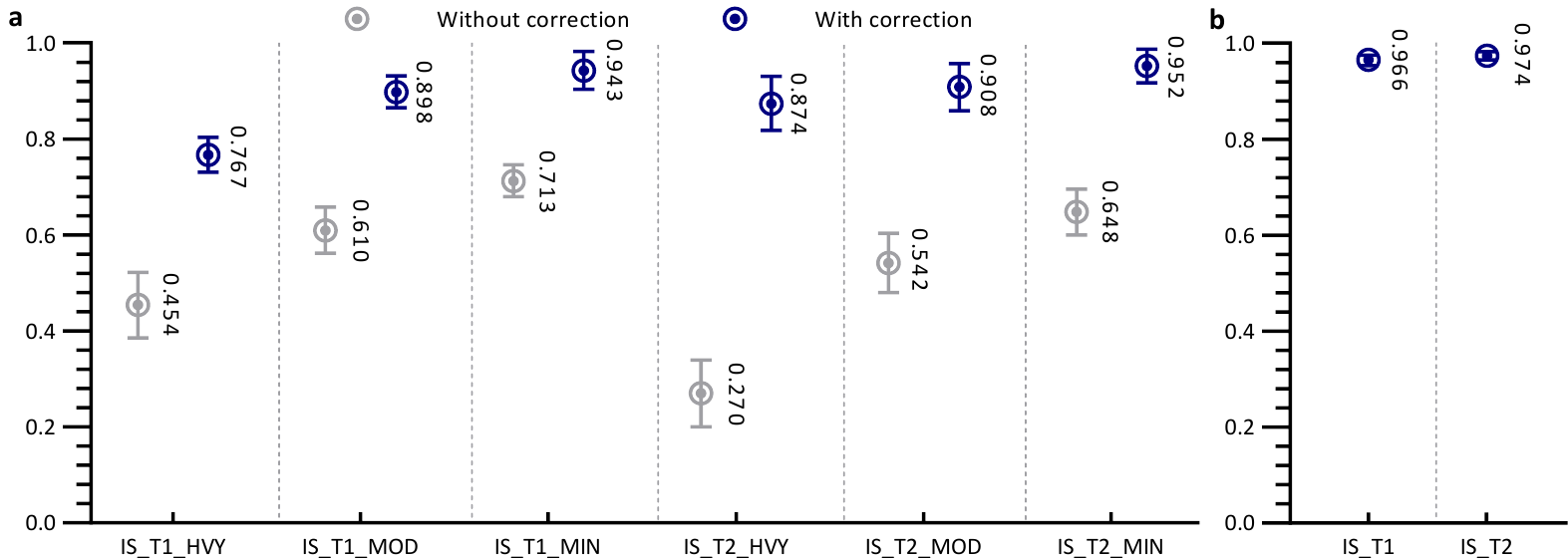}
	\caption{\textbf{Segmentation accuracy of in silico images.} 
    \textbf{a}, DSC comparison of artifact-corrupted images with and without correction, indicating artifact removal improves image usability.
	\textbf{b}, Applying DUNCAN on the artifact-free images do not degrade image details, as indicated by the high DSCs. The bars show the means and the error bars show the standard errors on the means. The sample size is 10 for each case.
}
	\label{fig:sim_seg_comp}
\end{figure*}

\subsection{Datasets}
We evaluated DUNCAN using 
\begin{inparaenum}[(i)]
  \item An in vivo dataset of T1- and T2-weighted images of children scanned from one month to six years of age\cite{Howell2019The}; and
  \item An in silico dataset of T1- and T2-weighted images with simulated artifacts.
\end{inparaenum}

We denote the in vivo datasets for T1- and T2-weighted MR images as IV\_T1 and IV\_T2, respectively. 
For each modality, we selected 20 artifact-free and 20 artifact-corrupted volumes for training, and 10 artifact-corrupted volumes for testing. 
We extracted 76 to 85 axial slices from each image volume, resulting in a total of 
1620, 1600, and 800 axial slices respectively from the 20 artifact-free, 20 artifact-corrupted, and 10 artifact-corrupted T1-weighted volumes for IV\_T1 and 
1520, 1550, and 800 axial slices respectively from the 20 artifact-free, 20 artifact-corrupted, and 10 artifact-corrupted T2-weighted volumes for IV\_T2.
Each artifact-corrupted image volume was labeled with one of three artifact severity levels: minor, moderate, or heavy. 

To generate the in silico datasets, we synthesized artifact-corrupted images from the artifact-free images from IV\_T1 and IV\_T2 with three levels of artifacts, i.e., minor, moderate, and heavy. The resulting datasets are respectively denoted as IS\_T1\_MIN, IS\_T1\_MOD, and IS\_T1\_HVY for T1-weighted images, and IS\_T2\_MIN, IS\_T2\_MOD, and IS\_T2\_HVY for T2-weighted images.
We simulated the motion artifacts in k-space, reflecting background noise movements, swallowing-like movements, and random sudden movements. We generated the background noise movement via a pseudorandomized series (Perlin noise\cite{Perlin1985An}) with magnitude of 5, the swallowing-like movements via multiplications with linear phase shifts in motion directions, i.e., translations along $z$-axis and rotations along $x$-axis, and the random sudden movements via sudden changes in the magnitudes of motions in all directions.
For IS\_T1\_MIN, IS\_T1\_MOD, and IS\_T1\_HVY, 1620, 1620, 800, and 800 axial slices were extracted, respectively, from the 20 artifact-free, 20 synthesized artifact-corrupted, 10 artifact-free, and 10 synthesized artifact-corrupted T1-weighted volumes.
For IS\_T2\_MIN, IS\_T2\_MOD, and IS\_T2\_HVY, 1520, 1520, 800, and 800 axial slices were extracted, respectively, from the 20 artifact-free, 20 synthesized artifact-corrupted, 10 artifact-free, and 10 synthesized artifact-corrupted T2-weighted volumes.

\subsection{Evaluation Metrics}
To quantitatively evaluate the performance of DUNCAN on the in silico datasets, 
several image quality metrics, including mean square error (MSE), structural similarity index (SSIM)\cite{Wang2004Image}, multi-scale structural similarity index (MS-SSIM)\cite{Wang2003Multiscale}, peak signal-to-noise ratio (PSNR), visual information fidelity (VIF)\cite{Sheikh2006Image}, and universal quality index (UQI)\cite{Wang2002A}, were utilized to gauge the quality of the artifact-corrected images. 
We used the default settings for all the hyper-parameters of the evaluation metrics. 
For all metrics, except MSE, higher values indicate better performance.

\subsection{Compared Methods}
To verify the effectiveness and superiority of DUNCAN, we compared it with three state-of-the-art methods that are closely related to our task, i.e., one supervised method -- U-Net\cite{Jin2017Deep} -- and two unsupervised methods -- CycleGAN\cite{Zhu2017Unpaired} and Pix2Pix\cite{Isola2017Image} -- implemented with Keras\textsuperscript{\ref{git:cycleGAN,Pix2Pix}}. Pix2Pix differs from CycleGAN by using a least-squares adversarial loss\cite{Mao2017Least} and PatchGAN\cite{Isola2017Image} as the discriminator. 
\footnotetext[1]{https://github.com/eriklindernoren/Keras-GAN/\label{git:cycleGAN,Pix2Pix}}

\subsection{Performance Evaluation Using In Vivo Datasets}
Since no ground truth is available for the in vivo images, only qualitative comparisons were conducted.
The comparison results for different levels of artifacts are shown for the T1-weighted and T2-weighted datasets in Figures~\ref{fig:CL_comp}a and \ref{fig:CL_comp}b, respectively.
CycleGAN and Pix2Pix are unable to remove the artifacts completely for different levels of artifacts in the T1- and T2-weighted images.
In comparison, DUNCAN is able to remove artifacts with varying severity without introducing new artifacts.

\subsection{Performance Evaluation Using In Silico Datasets}
Visual comparison results for the in silico T1- and T2-weighted datasets are provided in Figures~\ref{fig:SYN_comp_qual}a and \ref{fig:SYN_comp_qual}b, respectively. The error maps, gradient maps, and gradient error maps for T1- and T2-weighted images are  provided in Supplementary Figures~1--3, respectively. 
Quantitative comparison results using various evaluation metrics 
are summarized in Figures~\ref{fig:SYN_T1_comp_quan} and \ref{fig:SYN_T2_comp_quan}, respectively.
Quantitative comparison results of gradient maps using various evaluation metrics on in silico T1- and T2-weighted images are included in Supplementary Figure~4.
CycleGAN and Pix2Pix yield similar performance for the various evaluation metrics,
but in terms of visual appearance, Pix2Pix is significantly better than CycleGAN due to its PatchGAN discriminator and least-squares adversarial loss.
Although U-Net was trained with paired data and performs better than CycleGAN and Pix2Pix for the various evaluation metrics, CycleGAN and Pix2Pix generate images that are sharper than U-Net, both qualitatively and quantitatively, as illustrated in Supplementary Figures~2--4. 
This is due to the use of adversarial learning in CycleGAN and Pix2Pix.
In comparison, as DUNCAN utilizes both adversarial learning and disentangled representation learning of artifacts and contents, it yields better performance in artifact removal and better capability in maintaining structural information in artifact-corrupted images. 
Even when corrupted with heavy artifacts, image details can still be satisfactorily recovered by DUNCAN.

\subsection{Tissue Segmentation}
To further demonstrate that DUNCAN can improve image usability,  
we applied BET \cite{Smith2002Fast} and FAST \cite{Zhang2001Segmentation} on the testing data in IS\_T1 and IS\_T2 for brain extraction and tissue segmentation, respectively. 
We report in Figure~\ref{fig:sim_seg_comp} the tissue segmentation accuracy before and after artifact correction, as measured by the Dice similarity coefficient (DSC). Tissue segmentation maps from the artifact-free images were used as references. The results shown in Figure~\ref{fig:sim_seg_comp}a indicate that DSCs are  improved remarkably by DUNCAN correction. 
To validate the quality preservation property of DUNCAN for artifact-free images, we also evaluated the tissue segmentation accuracy of artifact-free images processed with DUNCAN.
The high DSCs shown in Figure~\ref{fig:sim_seg_comp}b indicate that the quality of artifact-free images is preserved.

\subsection{Discussion}
We have demonstrated that DUNCAN can be applied to MR images for post-acquisition artifact removal. DUNCAN is therefore useful when raw k-space data and reconstruction algorithms are not available. DUNCAN is a flexible method for RAC irrespective of the acquisition technique. For training, the user only needs to label the images as artifact-corrupted or artifact-free. No additional images need to be acquired and no knowledge of MR physics is needed to simulate artifacts.
DUNCAN can potentially allow image imperfections such as noise, streaking, and ghosting to be removed without explicitly generating them for supervised training. DUNCAN can be incorporated in a quality control pipeline to improve image usability. We also note that DUNCAN can be used, via translation of artifact-free to artifact-corrupted images, to generate natural and realistic artifacts that can be used for supervised or unsupervised training of machine learning algorithms.


\begin{addendum}
	\item[Acknowledgments] This work was supported in part by National Institutes of Health grants (EB006733, AG053867, MH117943, MH104324, MH110274) and the efforts of the UNC/UMN Baby Connectome Project Consortium. 
	The authors thank Dr.~Xiaopeng Zong of the University of North Carolina at Chapel Hill for an initial discussion on motion artifact simulation and Dr.~Yoonmi Hong of the University of North Carolina at Chapel Hill and Dr.~Yong Chen of Case Western Reserve University for proofreading the paper.
	\item[Competing Interests] The authors declare that they have no
	competing financial interests.
	\item[Correspondence] Correspondence and requests for materials
	should be addressed to Pew-Thian Yap~(email: ptyap@med.unc.edu).
	\item[Data Availability] The data used in this paper were provided by the investigative team of the  UNC/UMN Baby Connectome Project. The data can be obtained from the National Institute of Mental Health Data Archive (NDA) (\url{http://nda.nih.gov/}) or by contacting the investigative team\cite{Howell2019The}.
	\item[Code Availability] The source code and trained models for this study are publicly available on Zenodo (\url{https://zenodo.org/record/3742351})\cite{Liu2020Code}.
	\item[Author Contribution] SL designed the framework and network architecture, carried out the implementation, performed the experiments, and analyzed the data. SL and PTY wrote the manuscript.
  SL, KHT, and PTY revised the manuscript.
  LQ contributed to the initial formulation of the method before moving to Stanford University. WL provided the infant data for training and testing.
  PTY conceived the study and were in charge of overall direction and planning. DS was involved in the initial discussion of the problem when he was with the University of North Carolina at Chapel Hill. All work was done at the University of North Carolina at Chapel Hill.
\end{addendum}

\begin{methods}
In this work, we 
\begin{inparaenum}[(i)]
	\item consider the artifact removal problem as image translation from an artifact-corrupted domain to an artifact-free domain;
	\item propose an end-to-end unsupervised RAC framework based on a disentangled unsupervised cycle-consistent adversarial network (DUNCAN, see Figure~\ref{fig:framework}), which employs two auto-encoders to learn a cycle translation mapping that translates the images forward and backward between the artifact-corrupted and artifact-free domains;
	\item adopt two encoders to embed the images in a domain-invariant content space, which contains anatomical information, and a domain-specific artifact space, which captures artifact and noise information, and adopt a decoder to translate the images to a target domain using the encoded content and artifact features;
	\item realize content-artifact disentanglement, hinging on determining the domain-invariant content space using two strategies: a multi-scale content consistency (MS-CC) loss to keep content features consistent across domains and a content-swapping mechanism to ensure the domain-invariance of the content space; and
	\item introduce a quality preservation mechanism to ensure that no image details are removed.
\end{inparaenum}

\subsection{Disentangled Cycle Translation Mapping}
Let $\mathcal{I}_{\text{c}}$ and $\mathcal{I}_{\text{f}}$ be the domains of artifact-corrupted and artifact-free MR images, respectively. 
Our method aims to learn the nonlinear mappings between the two domains, i.e., $\mathcal{M}_{\text{c}\rightarrow \text{f}}:\mathcal{I}_{\text{c}}\rightarrow\mathcal{I}_{\text{f}}$ and $\mathcal{M}_{\text{f}\rightarrow \text{c}}:\mathcal{I}_{\text{f}}\rightarrow\mathcal{I}_{\text{c}}$, using unpaired training images.
In practice, each acquired MR image, even with good quality, may inevitably contain artifacts. Therefore, we assume that each MR image is a nonlinear combination of content and artifact components.
For two unpaired images $(x_{\text{c}}\in\mathcal{I}_{\text{c}},x_{\text{f}}\in\mathcal{I}_{\text{f}})$, disentangled cycle translation (DCT) mapping $\mathcal{M}_{\text{c}\rightarrow \text{f}\rightarrow \text{c}}:\mathcal{I}_{\text{c}}\rightarrow\mathcal{I}_{\text{f}}\rightarrow\mathcal{I}_{\text{c}}$ is accomplished with sequential forward and backward translation mappings $\mathcal{M}_{\text{c}\rightarrow \text{f}}$ and $\mathcal{M}_{\text{f}\rightarrow \text{c}}$; conversely, the DCT mapping $\mathcal{M}_{\text{f}\rightarrow \text{c}\rightarrow \text{f}}:\mathcal{I}_{\text{f}}\rightarrow\mathcal{I}_{\text{c}}\rightarrow\mathcal{I}_{\text{f}}$ is realized with $\mathcal{M}_{\text{f}\rightarrow \text{c}}$ and $\mathcal{M}_{\text{c}\rightarrow \text{f}}$, as illustrated in Figure~\ref{fig:framework}a. 
Specifically, taking DCT mapping $\mathcal{M}_{\text{c}\rightarrow \text{f}\rightarrow \text{c}}$ as an example, for forward translation mapping $\mathcal{M}_{\text{c}\rightarrow \text{f}}$, we first encode $x_{\text{c}}$ in two latent spaces, i.e., the domain-invariant content space $\mathcal{C}$ and the domain-specific artifact space $\mathcal{A}_{\text{c}}$, to obtain two disentangled representations, i.e., the artifact (AF) representation $z_{\text{c}}^{\text{AF}}$ and the content (CT) representation $z_{\text{c}}^{\text{CT}}$, respectively. 
We then build a decoder based on the disentangled representations to construct intermediate image $x_{\text{c}\rightarrow \text{f}}$ in the artifact-free domain $\mathcal{I}_{\text{f}}$. The forward translation mapping $\mathcal{M}_{\text{c}\rightarrow \text{f}}$ for $x_{\text{c}}$ can be summarized as $\{x_{\text{c}} \in \mathcal{I}_{\text{c}}\} \rightarrow \{z_{\text{c}}^{\text{CT}} \in \mathcal{C}, z_{\text{c}}^{\text{AF}} \in \mathcal{A}_{\text{c}}\} \rightarrow \{x_{\text{c}\rightarrow \text{f}} \in \mathcal{I}_{\text{f}}\}$.
We then further conduct a backward translation mapping $\mathcal{M}_{\text{f}\rightarrow \text{c}}$ on $x_{\text{c}\rightarrow \text{f}}$. We first encode $x_{\text{c}\rightarrow \text{f}}$ as $z_{\text{c}\rightarrow \text{f}}^{\text{CT}}$ in content space $\mathcal{C}$ and $z_{\text{c}\rightarrow \text{f}}^{\text{AF}}$ in artifact space $\mathcal{A}_{\text{f}}$. 
Note that this artifact space is specific to input from domain $\mathcal{I}_{\text{f}}$, whereas $\mathcal{A}_{\text{c}}$ is specific to input from domain $\mathcal{I}_{\text{c}}$, as images from both domains have different types of artifacts manifesting in different styles. 
Feeding $z_{\text{c}\rightarrow \text{f}}^{\text{CT}}$ and $z_{\text{c}\rightarrow \text{f}}^{\text{AF}}$ as input to a decoder, we obtain the reconstructed artifact-corrupted image $\hat{x}_{\text{c}}$. 
The backward translation mapping $\mathcal{M}_{\text{f}\rightarrow \text{c}}$ for $x_{\text{c}\rightarrow \text{f}}^{\text{CT}}$ can be summarized as $\{x_{\text{c}\rightarrow \text{f}}^{\text{CT}} \in \mathcal{I}_{\text{f}}\} 
\rightarrow \{z_{\text{c}\rightarrow \text{f}}^{\text{CT}} \in \mathcal{C}, z_{\text{c}\rightarrow \text{f}}^{\text{AF}} \in \mathcal{A}_{\text{f}}\} \rightarrow \{\hat{x}_{\text{c}} \in \mathcal{I}_{\text{c}}\}$. 
Similarly, we also perform DCT mapping for $x_{\text{f}}$ via $\mathcal{M}_{\text{f}\rightarrow \text{c}}$ and $\mathcal{M}_{\text{c}\rightarrow \text{f}}$ to obtain in sequence the intermediate artifact-corrupted image $x_{\text{f}\rightarrow \text{c}}$
 and artifact-free image $\hat{x}_{\text{f}}$, i.e., $\{x_{\text{f}} \in \mathcal{I}_{\text{f}}\} \rightarrow \{z_{\text{f}}^{\text{CT}} \in \mathcal{C}, z_{\text{f}}^{\text{AF}} \in \mathcal{A}_{\text{f}} \} \rightarrow \{x_{\text{f}\rightarrow \text{c}} \in \mathcal{I}_{\text{c}}\} \rightarrow \{z_{\text{f}\rightarrow \text{c}}^{\text{CT}} \in \mathcal{C}, z_{\text{f}\rightarrow \text{c}}^{\text{AF}} \in \mathcal{A}_{\text{c}} \} \rightarrow \{\hat{x}_{\text{f}} \in \mathcal{I}_{\text{f}}\}$.

\subsection{The DUNCAN Architecture}
Figure~\ref{fig:framework}b shows an overview of the network architecture of DUNCAN, consisting of two DCT mappings (artifact-corrupted and artifact-free domain cycles), two content-swapping and identity translations (in both artifact-corrupted and artifact-free domains), and four adversarial constraints
(for the generated artifact-corrupted and artifact-free images). 
As described in the previous section, the artifact-corrupted and artifact-free domain cycles aim to perform DCT mappings $\mathcal{M}_{\text{c}\rightarrow\text{f}\rightarrow\text{c}}$ and $\mathcal{M}_{\text{f}\rightarrow\text{c}\rightarrow\text{f}}$, respectively, using two domain translation mappings, i.e., $\mathcal{M}_{\text{f}\rightarrow\text{c}}$ and $\mathcal{M}_{\text{c}\rightarrow\text{f}}$. Each domain translation mapping is realized by two encoders to disentangle an image into content and artifact features, and one decoder to reconstruct the target-domain image using the disentangled features, as illustrated in Figure~\ref{fig:framework}c. More specifically, the mapping $\mathcal{M}_{\text{f}\rightarrow\text{c}}$ is realized by content encoder $E_{\text{f}}^{\text{CT}}$, artifact encoder $E_{\text{f}}^{\text{AF}}$, and decoder $D_{\text{f}}$, whereas the mapping $\mathcal{M}_{\text{c}\rightarrow \text{f}}$ is realized by content encoder $E_{\text{c}}^{\text{CT}}$, artifact encoder $E_{\text{c}}^{\text{AF}}$, and decoder $D_{\text{c}}$. With any two unpaired images $x_{\text{c}} \in \mathcal{I}_{\text{c}}$ and $x_{\text{f}} \in \mathcal{I}_{\text{f}}$ as inputs, the encoders and decoders are learned to respectively reconstruct images $\hat{x}_{\text{c}}$ and $\hat{x}_{\text{f}}$ via DCT mappings $\mathcal{M}_{\text{c}\rightarrow \text{f}\rightarrow \text{c}}$ and $\mathcal{M}_{\text{f}\rightarrow \text{c}\rightarrow \text{f}}$.

Using the domain-invariant property of the content space, we propose content-swapping translation (CST) for complete representation disentanglement, as illustrated in Figure~\ref{fig:framework}c.
The idea behind this mechanism is that when the content and artifact information are completely disentangled, swapping the domain-invariant content information between domain translations should not lead to 
changes in translation outcomes.
Specifically, we replace content information from $E_{\text{c}}^{\text{CT}}(x_{\text{c}})$ in domain translation $\mathcal{M}_{\text{c}\rightarrow \text{f}}$ with content information from $E_{\text{f}}^{\text{CT}}(x_{\text{f}})$ to construct content-swapped translated image $\hat{x}_{\text{c}\leftrightarrow \text{f}} \in \mathcal{I}_f$ via decoder $D_{\text{c}}$. Similarly, we replace content information from $E_{\text{f}}^{\text{CT}}(x_{\text{f}})$ in domain translation $\mathcal{M}_{\text{f}\rightarrow \text{c}}$ with content information from $E_{\text{c}}^{\text{CT}}(x_{\text{c}})$ to generate content-swapped translated image $\hat{x}_{\text{f}\leftrightarrow \text{c}} \in \mathcal{I}_{\text{c}}$ via decoder $D_{\text{f}}$. 
The translated images $\hat{x}_{\text{c}\leftrightarrow\text{f}}$ and $\hat{x}_{\text{f}\leftrightarrow\text{c}}$, respectively, are constrained by $x_{\text{f}}$ and $x_{\text{c}}$ via discriminators $D_{\text{c}\leftrightarrow\text{f}}^{\textrm{ADV}}$ and $D_{\text{f}\leftrightarrow\text{c}}^{\textrm{ADV}}$. 

Furthermore, we constrain the forward translation mappings with identity translation (IT) mappings, as illustrated in Figure~\ref{fig:framework}d, to maintain image quality when no alteration is expected.
Specifically, when translation mappings $\mathcal{M}_{\text{c}\rightarrow \text{f}}$ and $\mathcal{M}_{\text{f}\rightarrow \text{c}}$ are applied to $x_{\text{f}}$ and $x_{\text{c}}$, respectively, the mappings are constrained to result in identity reconstructions $\tilde{x}_{\text{f}} \in \mathcal{I}_{\text{f}}$ and $\tilde{x}_{\text{c}} \in \mathcal{I}_{\text{c}}$. A set of consistency losses are used to ensure that the identity translated images are consistent with the corresponding input images.

\subsection{Auto-Encoder Architecture}
We devise two auto-encoders $G_{\text{c}}$ and $G_{\text{f}}$ to respectively generate artifact-free and artifact-corrupted images.
Each auto-encoder, with architecture illustrated in Figure~\ref{fig:framework}e, consists of 
\begin{inparaenum}[(i)]
  \item 
a content encoder, $E_{\text{c}}^{\text{CT}}$ or $E_{\text{f}}^{\text{CT}}$, and an artifact encoder, $E_{\text{c}}^{\text{AF}}$ or $E_{\text{f}}^{\text{AF}}$, that respectively map an input image to the domain-invariant latent space $\mathcal{C}$ and the domain-specific latent space, $\mathcal{A}_{\text{c}}$ or $\mathcal{A}_{\text{f}}$, to respectively extract the content and artifact information of the image, and 
\item a decoder,  $\mathcal{D}_{\text{c}}$ or $\mathcal{D}_{\text{f}}$,
that generates from the extracted features an image in the target domain of the auto-encoder.
\end{inparaenum}
We describe next the details for each component of the auto-encoder.

\textbf{Content Encoder} The content encoder takes the original image as input, and extracts content features through 4 residual blocks. 
Each residual block consists of 4$\times$4 convolution, leaky ReLU, and instance normalization (IN)\cite{Ulyanov2017Improved} layers. We use an IN layer instead of a batch normalization layer\cite{Ulyanov2017Improved} to accelerate model convergence and maintain independence between features.
All normalized feature maps are activated by leaky ReLU with negative slope 0.2.

\textbf{Artifact Encoder} We first down-sample the input image using 2$\times$2 average-pooling in the artifact encoder. Then, we extract features from the down-sampled image using 3 residual blocks without IN layers since IN removes the feature means and variances, which contain important artifact information.

\textbf{Decoder} The decoder takes the extracted content and artifact features as input and generates, using a set of up-sampling layers and residual blocks, a content image and an artifact image, which are concatenated and then fused through a residual block and an 1$\times$1 convolution layer to generate the translated image. 

\subsection{Adversarial Learning}
We employ generative adversarial networks (GANs) to better learn the 
 translation mappings between the artifact-free and artifact-corrupted image domains. A 
GAN is comprised of a generator network and a discriminator network. In our case, the auto-encoder acts as the generator network by translating an input image to a target-domain image. The discriminator network is a classifier that distinguishes between real and fake images. As training progresses, the generator is getting better at fooling the discriminator, and the discriminator is getting better at distinguishing real and fake images. 
We employ two discriminators $D_{\text{c}}^{\textrm{ADV}}$ and $D_{\text{f}\leftrightarrow\text{c}}^{\textrm{ADV}}$ in the artifact-corrupted domain and another two discriminators $D_{\text{f}}^{\textrm{ADV}}$ and $D_{\text{c}\leftrightarrow\text{f}}^{\textrm{ADV}}$ in the artifact-free domain. 
We use PatchGAN\cite{Isola2017Image}, shown in Figure~\ref{fig:framework}f, as the discriminators. 
The numbers of filters are 64, 128, 256, 512 for the convolution layers and the number of output channel is 1. Leaky ReLU activation is implemented with a negative slope coefficient of 0.2.

\subsection{Disentangled Representation Learning}
We took several measures to ensure that our encoders can properly disentangle content and artifact information from an input image. First, as content space $\mathcal{C}$ is domain-invariant, i.e., shared between the artifact-free and artifact-corrupted domains, the content information of an image and its generated counterpart in the target domain should be consistent. For example, the content information of $x_{\text{c}}$ and $x_{\text{c}\rightarrow\text{f}}$ should be consistent, and so should $x_{\text{f}}$ and $x_{\text{f}\rightarrow\text{c}}$.  
To this end, we propose a multi-scale content consistency (MS-CC) loss based on the low- and high-level features of the content encoders to respectively encourage the consistency of structural and semantic content information.
Second, discriminating between real and content-swapped generated images via discriminators $D_{\text{f}\leftrightarrow\text{c}}^{\text{ADV}}$ and $D_{\text{c}\leftrightarrow\text{f}}^{\text{ADV}}$ also ensures better disentanglement by the encoders.

\subsection{Image Quality Consistency (IQC)}
To ensure that the image quality of an input image is similar to the translated image, we propose a pixel-wise image quality consistency (IQC) loss to encourage the auto-encoders in the translation mappings $\mathcal{M}_{\text{f}\rightarrow \text{c}}$ and $\mathcal{M}_{\text{c}\rightarrow \text{f}}$ to perform as identity translation mappings $\mathcal{M}_{\text{c}\rightarrow \text{c}}:\mathcal{I}_{\text{c}}\rightarrow\mathcal{I}_{\text{c}}$ and $\mathcal{M}_{\text{f}\rightarrow \text{f}}:\mathcal{I}_{\text{f}}\rightarrow\mathcal{I}_{\text{f}}$ when respectively given artifact-corrupted and artifact-free images.
The IQC loss encourages the artifact-free image generator to not remove image details when given a good quality image. Similarly, the IQC loss encourages the artifact-corrupted image generator to not introduce any additional artifacts when given an artifact-corrupted image.

\subsection{Loss Functions}
We leverage two types of loss functions, i.e., consistency losses and adversarial losses to facilitate model training, as illustrated in Figures~\ref{fig:framework}c and \ref{fig:framework}d.
 
\textbf{Consistency Losses}
We utilize three consistency losses: multi-scale content consistency (MS-CC) loss $\mathcal{L}_{\textrm{MS-CC}}$, which measures content consistency between the input and output of the forward translation of each DCT (i.e., $\mathcal{M}_{\text{c}\rightarrow \text{f}}$ in $\mathcal{M}_{\text{c}\rightarrow \text{f}\rightarrow \text{c}}$ and $\mathcal{M}_{\text{f}\rightarrow \text{c}}$ in $\mathcal{M}_{\text{f}\rightarrow \text{c}\rightarrow \text{f}}$),
multi-scale reconstruction consistency (MS-RC) loss $\mathcal{L}_{\textrm{MS-RC}}$, which computes the reconstruction consistency between an image and its reconstructed counterpart in the same domain, and image quality consistency (IQC) loss, which computes the image quality consistency between an image and its identity translated counterpart in the same domain.

The MS-CC loss measures low- and high-level content feature differences
and is formulated as
\begin{equation}
\mathcal{L}_{\textrm{MS-CC}}=\sum_i[\mathbb{E}_{x_{\text{c}}\sim\mathcal{I}_{\text{c}}}\|\phi^i_{\text{c}}(x_{\text{c}})-\phi^i_{\text{f}}(x_{\text{c}\rightarrow\text{f}})\|_1+\mathbb{E}_{x_{\text{f}}\sim\mathcal{I}_{\text{f}}}\|\phi^i_{\text{f}}(x_{\text{f}})-\phi^i_{\text{c}}(x_{\text{f}\rightarrow\text{c}})\|_1],
\end{equation}
where $x_{\text{c}}$ and $x_{\text{f}}$ denote the artifact-corrupted and artifact-free images, respectively, and $x_{\text{c}\rightarrow\text{f}}=D_{\text{c}}(z_{\text{c}}^{\text{CT}},z_{\text{c}}^{\text{AF}})$ and $x_{\text{f}\rightarrow\text{c}}=D_{\text{f}}(z_{\text{f}}^{\text{CT}},z_{\text{f}}^{\text{AF}})$ denote the corresponding artifact-free and artifact-corrupted images generated by the decoders, respectively.
$\phi^i_{\text{c}}(\cdot)$ and $\phi^i_{\text{f}}(\cdot)$ 
denote the outputs of the $i$-th residual block of the content encoders $E_{\text{c}}^{\text{CT}}$ and $E_{\text{f}}^{\text{CT}}$, respectively.
$z_{\text{c}}^{\text{CT}}=E_{\text{c}}^{\text{CT}}(x_{\text{c}})\in\mathcal{S}$ and $z_{\text{f}}^{\text{CT}}=E_{\text{f}}^{\text{CT}}(x_{\text{f}})\in\mathcal{S}$ denote the content information extracted respectively from $x_{\text{c}}$ and $x_{\text{f}}$, whereas $z_{\text{c}}^{\text{AF}}=E_{\text{c}}^{\text{AF}}(x_{\text{c}})\in\mathcal{A}_{\text{c}}$ and $z_{\text{f}}^{\text{AF}}=E_{\text{f}}^{\text{AF}}(x_{\text{f}})\in\mathcal{A}_{\text{f}}$ denote the artifact information extracted respectively from $x_{\text{c}}$ and $x_{\text{f}}$. 

We compute the MS-RC loss by combining three reconstruction consistency losses, i.e., the pixel reconstruction consistency (PRC) loss $\mathcal{L}_{\textrm{PRC}}$, the edge reconstruction consistency (ERC) loss $\mathcal{L}_{\textrm{ERC}}$, and the structure reconstruction consistency (SRC) loss $\mathcal{L}_{\textrm{SRC}}$, as defined below:
\begin{align}
\mathcal{L}_{\textrm{MS-RC}}&=\mathcal{L}_{\textrm{PRC}}+\mathcal{L}_{\textrm{ERC}}+\mathcal{L}_{\textrm{SRC}},\\
\mathcal{L}_{\textrm{PRC}} &= 
\mathbb{E}_{x_{\text{c}}\sim\mathcal{I}_{\text{c}}}
\|x_{\text{c}}-\hat{x}_{\text{c}}\|_1
+
\mathbb{E}_{x_{\text{f}}\sim\mathcal{I}_{\text{f}}}\|x_{\text{f}}-\hat{x}_{\text{f}}\|_1,\\
\mathcal{L}_{\textrm{ERC}} &= \mathbb{E}_{x_{\text{c}}\sim\mathcal{I}_{\text{c}}}\|\psi_{\text{L}}(x_{\text{c}})-\psi_{\text{L}}(\hat{x}_{\text{c}})\|_1+\mathbb{E}_{x_{\text{f}}\sim\mathcal{I}_{\text{f}}}\|\psi_{\text{L}}(x_{\text{f}})-\psi_{\text{L}}(\hat{x}_{\text{f}})\|_1,\\
\mathcal{L}_{\textrm{SRC}} &= \mathbb{E}_{x_{\text{c}}\sim\mathcal{I}_{\text{c}}}
\|\psi_{\text{H}}(x_{\text{c}})-\psi_{\text{H}}(\hat{x}_{\text{c}})\|_1+
\mathbb{E}_{x_{\text{f}}\sim\mathcal{I}_{\text{f}}}\|\psi_{\text{H}}(x_{\text{f}})-\psi_{\text{H}}(\hat{x}_{\text{f}})\|_1,
\end{align}
where $\hat{x}_{\text{c}}=D_{\text{f}}(E_{\text{f}}^\text{CT}(x_{\text{c}\rightarrow\text{f}}),E_{\text{f}}^\text{AF}(x_{\text{c}\rightarrow\text{f}}))$ and $\hat{x}_{\text{f}}=D_{\text{c}}(E_{\text{c}}^\text{CT}(x_{\text{f}\rightarrow\text{c}}),E_{\text{c}}^\text{AF}(x_{\text{f}\rightarrow\text{c}}))$ are, respectively, the reconstructed images in the artifact-corrupted and artifact-free domains, and $\psi_{\text{L}}(\cdot)$ and $\psi_{\text{H}}(\cdot)$ are, respectively, the low-level structural information that reflects edges and high-level semantic information that reflects contents measured by a pre-trained network (i.e., VGG19\cite{Simonyan2015Very} trained on ImageNet).

To preserve 
image quality 
after translation when no alteration is expected, an image quality consistency (IQC) loss is devised to measure the pixel-wise difference between the input and identity mapped images as 
\begin{equation}
\mathcal{L}_{_\textrm{IQC}}=\mathbb{E}_{x_\text{c}\sim\mathcal{I}_\text{c}}\|x_\text{c}-\tilde{x}_\text{c}\|_1+\mathbb{E}_{x_\text{f}\sim\mathcal{I}_\text{f}}\|x_\text{f}-\tilde{x}_\text{f}\|_1,
\end{equation}
where $\tilde{x}_\text{c}=D_\text{f}(E_\text{f}^\text{CT}(x_\text{c}), E_\text{f}^\text{AF}(x_\text{c}))$ and $\tilde{x}_\text{f}=D_\text{c}(E_\text{c}^\text{CT}(x_\text{f}),E_\text{c}^\text{AF}(x_\text{f}))$ are the identity mapped images of $x_\text{c}$ and $x_\text{f}$, respectively.

\textbf{Adversarial Losses}
We apply two types of adversarial losses, i.e., single-domain adversarial (SD-ADV) loss and cross-domain adversarial (CD-ADV) loss, to enhance the judgment accuracy of the discriminators. 
All adversarial losses are designed with the mean square error function. 
The SD-ADV loss is calculated in a specific domain, i.e., $\mathcal{I}_\text{c}$ or $\mathcal{I}_\text{f}$, as
\begin{equation}
\begin{split}
\mathcal{L}_{_\textrm{SD-ADV}} =&\frac{1}{2}\mathbb{E}_{x_\text{c}\sim\mathcal{I}_\text{c}}(D^\textrm{ADV}_\text{c}(x_\text{c})-I)^2+\frac{1}{2}\mathbb{E}_{x_\text{f}\sim\mathcal{I}_\text{f}}(D^\textrm{ADV}_\text{c}(x_{\text{f}\rightarrow\text{c}}))^2\\
+&\frac{1}{2}\mathbb{E}_{x_\text{f}\sim\mathcal{I}_\text{f}}(D^\text{ADV}_\text{f}(x_\text{f})-I)^2+\frac{1}{2}\mathbb{E}_{x_\text{c}\sim\mathcal{I}_\text{c}}(D^\text{ADV}_\text{f}(x_{\text{c}\rightarrow\text{f}}))^2,
\end{split}
\end{equation}
where $D^\text{ADV}_\text{c}$ and $D^\text{ADV}_\text{f}$ are the discriminators used to distinguish between real and fake images respectively in domains $\mathcal{I}_\text{c}$ and $\mathcal{I}_\text{f}$, $I$ is a matrix of ones with size $N_1\times N_2$ matching the output of the discriminator. 
The cross-domain adversarial (CD-ADV) loss is defined as
\begin{equation}
\begin{split}
\mathcal{L}_{_\textrm{CD-ADV}}=& \frac{1}{2}\mathbb{E}_{x_\text{c}\sim\mathcal{I}_\text{c}}(D^\text{ADV}_{\text{f}\leftrightarrow\text{c}}(x_\text{c})-I)^2+\frac{1}{2}\mathbb{E}_{x_\text{f}\sim\mathcal{I}_\text{f},x_\text{c}\sim\mathcal{I}_\text{c}}(D^\text{ADV}_{\text{f}\leftrightarrow\text{c}}(\hat{x}_{\text{f}\leftrightarrow\text{c}}))^2\\
+&\frac{1}{2}\mathbb{E}_{x_\text{f}\sim\mathcal{I}_\text{f}}(D^\text{ADV}_{\text{c}\leftrightarrow\text{f}}(x_\text{f})-I)^2+\frac{1}{2}\mathbb{E}_{x_\text{c}\sim\mathcal{I}_\text{c},x_\text{f}\sim\mathcal{I}_\text{f}}(D^\text{ADV}_{\text{c}\leftrightarrow\text{f}}(\hat{x}_{\text{c}\leftrightarrow\text{f}}))^2,
\end{split}
\end{equation}
where $\hat{x}_{\text{c}\leftrightarrow\text{f}} = D_\text{c}(z_\text{f}^\text{CT},z_\text{c}^\text{AF})$ and $\hat{x}_{\text{f}\leftrightarrow\text{c}} = D_\text{f}(z_\text{c}^\text{CT}, z_\text{f}^\text{AF})$ are the images reconstructed by cross-domain content information, i.e., $z_\text{f}^\text{CT}$ or $z_\text{c}^\text{CT}$, and current-domain artifact information, i.e., $z_\text{c}^\text{AF}$ or $z_\text{f}^\text{AF}$.

\textbf{Total Loss} In summary, the total loss function of DUNCAN is
\begin{equation}\label{eq:TotalLoss}
\mathcal{L}_{\textrm{total}}=\mathcal{L}_{_\textrm{SD-ADV}}+\mathcal{L}_{_\textrm{CD-ADV}}+\lambda_{_\textrm{MS-CC}}\mathcal{L}_{_\textrm{MS-CC}}+\lambda_{_\textrm{PRC}}\mathcal{L}_{_\textrm{PRC}}+\lambda_{_\textrm{ERC}}\mathcal{L}_{_\textrm{ERC}}+\lambda_{_\textrm{SRC}}\mathcal{L}_{_\textrm{SRC}}+\lambda_{_\textrm{IQC}}\mathcal{L}_{_\textrm{IQC}},
\end{equation}
where $\lambda_{_\textrm{MS-CC}}$, $\lambda_{_\textrm{PRC}}$, $\lambda_{_\textrm{ERC}}$, $\lambda_{_\textrm{SRC}}$, and $\lambda_{_\textrm{IQC}}$ are the loss weights used for controlling the contributions of the terms in term in Equation~\eqref{eq:TotalLoss}.
\end{methods}

\subsection{Implementation Details} 
DUNCAN was implemented using Keras with Tensorflow backend. Evaluation was based on a machine with a CPU (Intel i7-8700K) and a GPU (NVIDIA GeForce GTX 1080Ti 11GB RAM). The Adam optimizer with $1\times10^{-4}$ learning rate was utilized for minimizing the loss function. For in vivo T1- and T2-weighted datasets, i.e., IV\_{T1} and IV\_{T2}, we used $\lambda_{_\textrm{MS-CC}}=5$, $\lambda_{_\textrm{PCC}}=10$, $\lambda_{_\textrm{ERC}}=5$, $\lambda_{_\textrm{SRC}}=5$, and $\lambda_{_\textrm{IQC}}=1$ for MS-CC, PRC, ERC, SRC, and IQC losses, respectively. For in silico T1- and T2-weighted datasets, i.e., IS\_{T1} and IS\_{T2}, we used $\lambda_{_\textrm{MS-SCC}}=10$, $\lambda_{_\textrm{PCC}}=20$, $\lambda_{_\textrm{ERC}}=10$, $\lambda_{_\textrm{SRC}}=10$, and $\lambda_{_\textrm{IQC}}=5$ for MS-CC, PRC, ERC, SRC, and IQC losses, respectively.
For both the in vivo and in silico datasets, every three adjacent slices in each volume were inserted into RGB channels of a color image,
which was then normalized to have a range between -1 and 1 and cropped to 208$\times$256 from the geometric center.
During training, one artifact-corrupted image and one artifact-free image were randomly selected each time as inputs.

\newpage
\subsection{References}

\end{document}